\documentclass[aps, prd, twocolumn, showpacs, showkeys, superscriptaddress]{revtex4-1}
\usepackage{graphicx}
\usepackage{bm}
\usepackage{amsmath}
\usepackage{amsfonts}
\usepackage{amssymb}

\begin{document}

\title{Nucleon structure functions at small $x$ via the Pomeron exchange in AdS space with a soft infrared wall}

\author{Akira~Watanabe}

\email{watanabe@phys.sinica.edu.tw}

\affiliation{Institute of Physics, Academia Sinica, Taipei 11529, Taiwan, Republic of China}

\author{Katsuhiko~Suzuki}

\email{katsu\_s@rs.kagu.tus.ac.jp}

\affiliation{Department of Physics, Tokyo University of Science, Shinjuku, Tokyo 162-8601, Japan}

\date{\today}

\begin{abstract}
We present analyses on nucleon structure functions at small Bjorken-$x$ in the framework of holographic QCD.
In this study, we improve the description of the target nucleon in the current setup of the holographic model by introducing a soft-wall AdS/QCD model, in which the AdS geometry is smoothly cut off at the infrared boundary.
Combining the improved Pomeron-nucleon coupling and the wave function of the 5D U(1) vector field with the BPST Pomeron exchange kernel, we obtain the structure functions.
Here we focus on the nonperturbative kinematical region, where $10^{-6} \leq x \leq 10^{-2}$ 
and $Q^2 \le$ a few~[GeV$^2$], and show that our calculations for $F_2^p$ and $F_L^p$ are consistent with experimental data of the deep inelastic scattering at HERA.
Furthermore, we find that the resulting longitudinal-to-transverse ratio of the structure functions, $F_L^p/F_T^p$, depends on both $x$ and $Q^2$.
\end{abstract}



\maketitle

\section{\label{sec:level1}Introduction}
Understanding the quark-gluon structure of the nucleon is one of the most important unsolved problems in high energy physics.
To reveal it, the deep inelastic lepton-nucleon scattering (DIS) has played a pivotal role over several decades.
The differential cross section of unpolarized DIS is expressed with the two independent structure functions $F_2 (x,Q^2)$ and $F_L (x,Q^2)$, where $x$ is the Bjorken scaling variable and $Q^2$ is the photon four-momentum squared. 
Since the structure functions are basically nonperturbative physical quantities, we cannot directly calculate them.
Moreover, particularly in the small $x$ region, gluons are the dominant ``constituents'' and the analyses become much more complicated.

In the circumstance, the birth of the anti-de Sitter/conformal field theory (AdS/CFT) correspondence (or gauge/string correspondence in general)~\cite{Maldacena:1997re,Gubser:1998bc,Witten:1998qj} was one of the breakthroughs in high energy physics, which has provided us with opportunities to analytically investigate the nonperturbative regions in gauge theories including quantum chromodynamics (QCD).
Using this correspondence, one can perform analyses of the strong coupling gauge theory in the usual Minkowski space by the classical theory of gravitation in the higher dimensional curved space.
So far, two ways have been proposed to apply the conjectured correspondence to QCD, the string theoretical top-down approach~\cite{Kruczenski:2003be,Kruczenski:2003uq,Sakai:2004cn,Sakai:2005yt} and the phenomenological bottom-up approach~\cite{Son:2003et,Erlich:2005qh,DaRold:2005zs}, according to the AdS/CFT dictionary.
For both of them, various applications to phenomenological studies were done successfully~\cite{Polchinski:2001tt,Shifman:2005zn,Erdmenger:2007cm,Erlich:2009me,DaRold:2005vr,BoschiFilho:2005yh,Brodsky:2006uqa,Forkel:2007ru,Pomarol:2008aa,Brodsky:2007hb,Grigoryan:2007vg,Panico:2008it,Grigoryan:2007wn,Kwee:2007dd,Krikun:2008tf,BallonBayona:2007rs,Pire:2008zf,Katz:2007tf,Kim:2008hx}.

The first study for DIS via AdS/CFT correspondence was done by Polchinski and Strassler~\cite{Polchinski:2002jw}, and the Pomeron exchange kernel was proposed by Brower, Polchinski, Strassler, and Tan (BPST)~\cite{Brower:2006ea}.  
The Pomeron was originally introduced to describe the high energy scattering in the 1960's based on analyticity of the $S$-matrix, and now it is assumed to be the complicated multi-gluon exchange in QCD.
Assuming the single-Pomeron exchange picture, one can express the $F_2$ structure function at small $x$ as
\begin{equation}
F_2 (x,Q^2) \propto x^{1-\alpha _0}, \label{eq:F2withPomeron}
\end{equation}
where $\alpha _0$ is the Pomeron intercept.
In their study, the Pomeron in the usual Minkowski space is identified as the reggeized graviton in the AdS space, and the BPST kernel describes the Pomeron (graviton) exchange contribution to the total cross section.
Subsequent to several elaborated analyses~\cite{Cornalba:2007zb,Brower:2007qh,Brower:2007xg,Cornalba:2008sp,Cornalba:2009ax,Cornalba:2010vk}, in Ref.~\cite{Brower:2010wf} the authors showed that the experimental data of the nucleon $F_2$ structure function at small $x$ and the scale dependence of the Pomeron intercept can be well reproduced by using the BPST kernel with the ``super local approximation'' for the Pomeron couplings.
Subsequently in Ref.~\cite{Watanabe:2012uc}, the present authors proposed another description of the structure functions of hadrons at small $x$ without assuming the super local approximation.
In the study, they considered the more phenomenologically realistic Pomeron couplings and showed that the consistent results for the nucleon $F_2$ can be obtained and structure functions for other hadrons can also be calculated in their model.

In this study, we improve the description of the target nucleon in the model setup of the previous study, and present more consistent analyses focusing on the highly nonperturbative kinematical region where $10^{-6} \leq x \leq 10^{-2}$ and $Q^2 \leq$ a few~[GeV$^2$].  
To realize the improvement, we describe the nucleon by introducing a soft-wall AdS/QCD model, in which the AdS geometry is smoothly cut off in the infrared (IR) region, instead of the hard-wall model adopted in the previous work~\cite{Watanabe:2012uc}.  
The hard-wall model, in which the sharp cutoff for the geometry is imposed, has often been used to describe hadronic properties,  because the calculations are much simpler and the analytical expressions can be easily obtained in general.  
However, when we consider masses of the excited mesons (not the ground state), there is a crucial difference between results obtained by the soft- and hard-wall  models.
In hard-wall models, the squared masses of mesons $m_n^2$ with the excitation number $n$ grow as $n^2$, while the correct Regge behavior, $m_n^2 \propto n$, can be reproduced only in the soft-wall model~\cite{Karch:2006pv}.
When we consider the nucleon, the situation becomes more complicated.
Nevertheless, it has been shown that for the soft-wall model, the correct Regge behavior can be reproduced~\cite{Abidin:2009hr}.  
Hence, it is not clear whether our previous treatment with the hard-wall model 
is adequate to the description of the structure function of the nucleon.   
To draw a conclusion for this question,
testing the soft-wall version of the previous study is needed, and this is a primary aim of this paper.  
In actual calculations, we improve the overlap function of the nucleon, which represents the nucleon-Pomeron coupling, within the soft-wall model of AdS/QCD.  
Moreover, we also investigate the BPST Pomeron kernel with the modification from the soft-wall model.

Analyses on the nucleon longitudinal structure function $F_L$ are also newly presented here.
In the quark-parton model, the structure functions are expressed as ${F_2} = x\sum\limits_q {e_q^2q\left( x \right)}$ and $F_L = 0$, respectively, where $e_q$ is the quark charge and $q(x)$ is the quark distribution function.
However, experimental data show that $F_L$ does not vanish, which is due to the complicated gluon interactions.
Hence, studying the longitudinal structure function provides a unique opportunity to investigate the nonperturbative effects in QCD.
In our framework, once we fix three adjustable parameters with the experimental data 
for $F_2^p$, we can predict the $F_L$ structure function without further ambiguities.
We also consider the longitudinal-to-transverse ratio of the structure functions, $R = F_L^p(x,Q^2)/F_T^p(x,Q^2)$, which is a completely non-trivial physical quantity especially in the kinematical region we focus on in this paper.

This paper is organized as follows.
In Sec.~\ref{sec:level2}, we present the whole setup of the model by introducing the BPST Pomeron exchange kernel .
We show how to obtain the Pomeron-nucleon coupling in the AdS space in Sec.~\ref{sec:level3}.
Subsequently in Sec.~\ref{sec:level4}, we present our numerical results for the nucleon structure functions at small $x$.
Finally Sec.~\ref{sec:level5} is devoted to the summary and discussion.

\section{\label{sec:level2}Holographic description of structure functions}
Following the previous studies~\cite{Brower:2010wf,Watanabe:2012uc}, the holographic description of structure functions can be realized with the BPST Pomeron exchange kernel and overlap functions (or form factors in the 5D AdS space).
The amplitude for high energy two-body scattering, $1+2 \to 3+4$, can be expressed with the two-dimensional impact parameter $\bm{b}$ and the fifth (or bulk) coordinates $z$ (for the incident particle) and $z'$ (for the target particle) in the eikonal representation as
\begin{align}
{\cal A} (s,t) = &2is \int d^2b e^{i \bm{k_\perp } \cdot \bm{b}} \int dzdz' \nonumber \\
&\times P_{13}(z) P_{24}(z') \{ 1-e^{i \chi (s,\bm{b},z,z')} \}, \label{eq:amplitude}
\end{align}
where $P_{13}(z)$ and $P_{24}(z')$ are overlap functions, which contain the information on internal structure of incoming and outgoing particles.

Here, we focus on the nucleon DIS, where $1=3=\gamma^*$ and $2=4=N$.  
In this case, the overlap functions are understood as density distributions in the AdS space.
The nucleon is an on-shell state, hence the nucleon overlap function $P_{24}(z')$ satisfies the normalization condition,
\begin{equation}
\int dz' P_{24}(z') = 1.  
\end{equation}

We calculate the structure function considering the $\gamma^* N$ forward scattering amplitude by virtue of the optical theorem.
Keeping only the leading contribution to the eikonal approximation in Eq.~\eqref{eq:amplitude}, i.e. considering the single-Pomeron exchange, the structure functions at small $x$ are written as
\begin{align}
F_{i}(x,Q^2)= &\frac{Q^2}{2 \pi ^2} \int d^2b \int dzdz' \nonumber \\
&\times P_{13}^{(i)}(z,Q^2) P_{24}(z') \mbox{Im} \chi (s,\bm{b},z,z'), \label{eq:Fi}
\end{align}
where $i=L$ is for the longitudinal  structure function $F_L$ and $i=2$ for $F_2$.

In the conformal limit, the analytical form of the imaginary part of $\chi (s,\bm{b},z,z')$ can be obtained, and the impact parameter integration can be demonstrated explicitly.
Hence Eq.~\eqref{eq:Fi} is rewritten as
\begin{align}
F_i &(x,Q^2) = \frac{g_0^2 \rho^{3/2}Q^2}{32 \pi ^{5/2}} \int dzdz' P_{13}^{(i)} (z,Q^2) P_{24}(z') \nonumber \\
&\hspace{39mm} \times (zz') \mbox{Im} [\chi_{c}(s,z,z')], \label{eq:Fic} \\
&\mbox{Im} [\chi_c(s,z,z') ] \equiv e^{(1-\rho)\tau} e^ {-(({\log ^2 z/z'})/{\rho \tau})} / {\tau^{1/2}}, \label{eq:kernel_c}
\end{align}
where
\begin{equation}
\tau =\log (\rho zz' s/2),
\end{equation}
and $g_0^2$ and $\rho$ are adjustable parameters of the model.

As mentioned above, the conformal kernel Eq.~\eqref{eq:kernel_c} is calculated in the conformal limit.
However, the conformal symmetry is certainly broken in real QCD due to the nonperturbative effect such as color confinement.
Since we concentrate on the nonperturbative soft kinematical region in this study, we also consider a modified kernel to break the conformal symmetry phenomenologically~\cite{Brower:2010wf}, 
\begin{align}
&\mbox{Im} [\chi_{mod} (s,z,z')] \equiv 
\mbox{Im} [\chi_c (s,z,z') ] \nonumber \\
&\hspace{30mm} + \mathcal{F} (s,z,z') \mbox{Im} [\chi_c(s,z,z_0^2/z') ],\label{eq:kernel_mod} \\
&\mathcal{F} (s,z,z') = 1 - 2 \sqrt{\rho \pi \tau} e^{\eta^2} \mbox{erfc}( \eta ), \nonumber \\
&\eta = \left( -\log \frac{zz'}{z_0^2} + \rho \tau \right) / {\sqrt{\rho \tau}}. \nonumber
\end{align}
The first term in Eq.~\eqref{eq:kernel_mod} is exactly the same as the conformal kernel, and the second term contains a smooth function with a parameter $z_0$, which somehow mimics the confinement effect in QCD~\footnote{This kernel is called hard-wall kernel in Ref.~\cite{Brower:2010wf}.
In this work the adjustable parameter $z_0$ is not a sharp cutoff of the AdS geometry but controls the size of the conformal breaking effects, in contrast to the treatment in Ref.~\cite{Watanabe:2012uc}}.
From the results in Refs.~\cite{Brower:2010wf,Watanabe:2012uc}, one can see that the added term is needed to reproduce the rapid rise of the experimental data for $F_2^p$ with $Q^2$ increasing.

\begin{figure}[bt!]
\includegraphics[width=80mm]{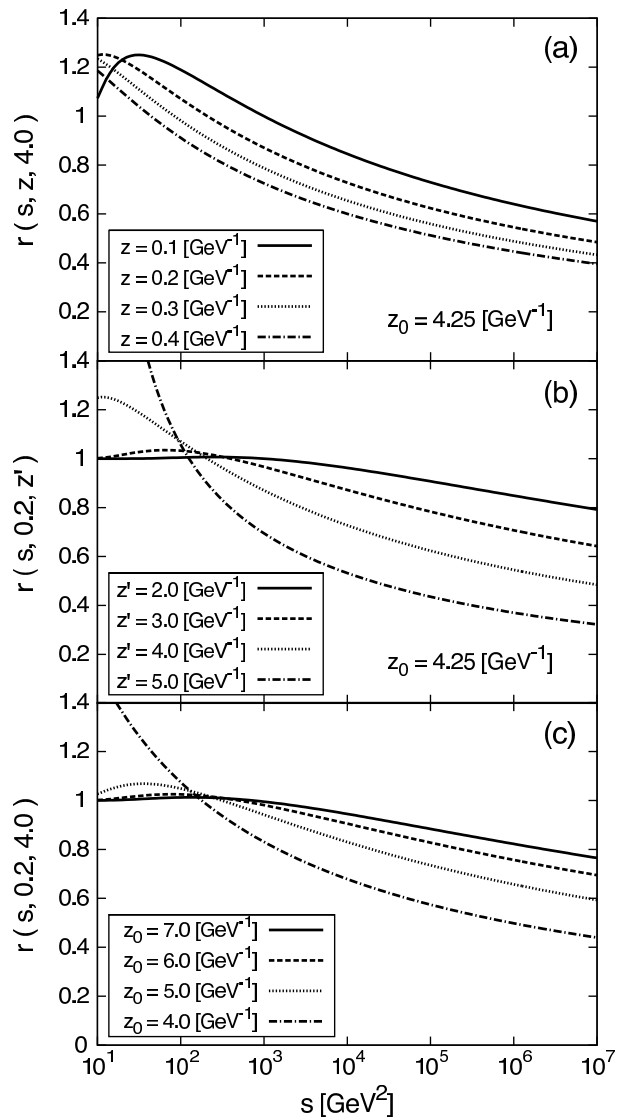}
\caption{
The kernel ratio $r(s,z,z')$ as a function of $s$.
The panels (a), (b), and (c) show the $z$, $z'$, and $z_0$ dependence of the ratio, respectively, at the values as follows.
(a) $z' = 4.0$~[GeV$^{-1}$], $z_0 = 4.25$~[GeV$^{-1}$].
(b) $z = 0.2$~[GeV$^{-1}$], $z_0 = 4.25$~[GeV$^{-1}$].
(c) $z = 0.2$~[GeV$^{-1}$], $z' = 4.0$~[GeV$^{-1}$].
The parameter $\rho$ is fixed at $\rho = 0.8$.
}
\label{fig:r_vs_s}
\end{figure}

\begin{figure}[bt!]
\includegraphics[width=80mm]{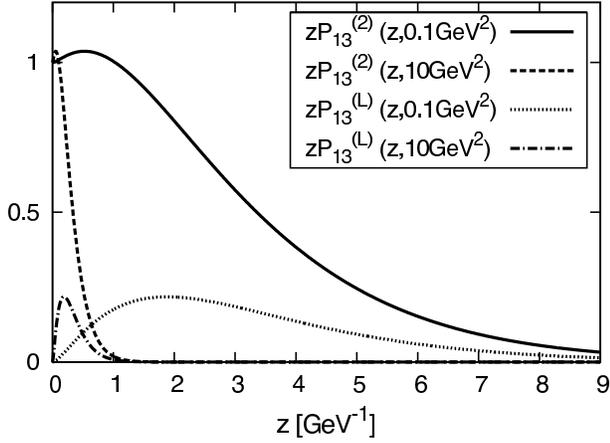}
\caption{
Behavior of the overlap functions for the virtual photon in the $z$ space.
The solid and dashed curves indicate $zP_{13}^{(2)}(z,Q^2)$ with $Q^2 = 0.1$~[GeV$^2$] and  $10$~[GeV$^2$], respectively.
The dotted and dash-dotted curves show $zP_{13}^{(L)}(z,Q^2)$ with $Q^2 = 0.1$~[GeV$^2$] and  $10$~[GeV$^2$], respectively.
}
\label{fig:OF13}
\end{figure}

To see the effect of the modified term more visually, we show in Fig.~\ref{fig:r_vs_s} the $s$ dependence of a ratio of the modified kernel to the conformal one defined by
\begin{equation}
r (s,z,z') \equiv \mbox{Im} [\chi_{mod} ] / \mbox{Im} [\chi_c ], \label{eq:kernel_ratio}
\end{equation}
for several values of the bulk coordinates $z$, $z'$ and a parameter $z_0$ (see caption in each panel).
As we will see in the next section (see Figs.~\ref{fig:OF13} and~\ref{fig:OF24}), the virtual photon overlap function $P_{13}(z)$ is localized  for relatively lower $z$, $z < 1$~[GeV$^{-1}$], while the nucleon counterpart is significant around $z' = 4.5$~[GeV$^{-1}$].
Thus, we take kinematical variables $0.1 \le z \le 0.4$ and $2.0 \le z' \le 5.0$ in Figs.~\ref{fig:r_vs_s}~(a) and (b) to clarify the effects of the modified kernel with an optimized parameter $z_0 = 4.25$~[GeV$^{-1}$].

From Fig.~\ref{fig:r_vs_s} we easily understand that the added term in the modified kernel Eq.~\eqref{eq:kernel_mod} gives a negative contribution, and thus the ratio becomes lower than 1 for large $s \geq 10^2$~[GeV$^2$].
The added term, however, enhances the kernel for lower $s$.
The results in Fig.~\ref{fig:r_vs_s}~(a) show that the reduction of the ratio due to the conformal breaking term  becomes larger with $z$ increasing, although the effect is not drastic.
On the other hand, it is also clear that the strong $s$ dependence of the modified kernel can be seen only if we take $z'\geq  4.5$~[GeV$^{-1}$] in Fig.~\ref{fig:r_vs_s}~(b).
This may indicate that the proper treatment of the nucleon  overlap function at larger $z'$ is important to describe the $s$ dependence of the structure function.
(Note that $z'$ is the bulk coordinate for the nucleon.)
For the choice of the parameter $z_0$, we show $r (s,z,z')$ for several values of $z_0$ in Fig.~\ref{fig:r_vs_s}~(c).
We obtain $z_0 = 4.25$~[GeV$^{-1}$] to  reproduce the data as we will show in Sec.~\ref{sec:level4}.

%

\section{\label{sec:level3}Pomeron couplings and a soft-wall AdS/QCD model}
To evaluate the structure functions expressed by Eq.~\eqref{eq:Fic}, we need to specify the overlap functions $P_{13}^{(i)} (z)$ and $P_{24} (z')$.  
First we consider $P_{13}^{(i)} (z)$ with $i=L$ and $i=2$.  $P_{13}^{(L)} (z)$ describes the longitudinally polarized photon, while $P_{13}^{(2)} (z)$ contains both transverse and longitudinal components.
Following the previous studies~\cite{Brower:2010wf,Watanabe:2012uc}, we adopt the massless 5D U(1) vector field, which was originally considered  by Polchinski and Strassler~\cite{Polchinski:2002jw}, to describe the off-shell state in the AdS space.
$P_{13}^{(i)} (z)$ are expressed as
\begin{align}
&P_{13}^{(2)}(z,Q^2) = Q^2 z \left( K_0^2(Qz) + K_1^2(Qz) \right), \label{eq:P132} \\
&P_{13}^{(L)}(z,Q^2) = Q^2 z K_0^2(Qz), \label{eq:P13L}
\end{align}
where $K_0$ and $K_1$ are the modified Bessel functions of the second kind.  
We show in Fig.~2 $zP_{13}^{(i)}(z,Q^2)$, which appeared in Eq.~(5), as functions of $z$ for $Q^2 = 0.1$~[GeV$^2$] and $10$~[GeV$^2$].  
Both $zP_{13}^{(L)}$ and $zP_{13}^{(2)}$ are localized near the origin for the larger  $Q^2$, while they penetrate into the large $z$ region for the smaller $Q^2$ case.  
This behavior is consistent with a naive expectation that the bulk coordinate $z$ gives a resolution scale of the virtual photon, $z \sim 1/Q$.
We have already found in Fig.~\ref{fig:r_vs_s}~(a) that  the nonperturbative conformal breaking contribution to the kernel is relevant for the larger $z$ case, which is nothing but the small $Q$ region.
As a result, the nonperturbative effect at small $Q^2$ is encoded in our calculation through the resolution scale $z$ provided by the virtual photon overlap function.

Next we consider the overlap function for the nucleon $P_{24}(z')$.
The Pomeron is assumed to be identified  with the reggeized graviton exchange in the AdS space.
Hence, we consider the Pomeron (graviton)-nucleon-nucleon three point function in the 5D classical action of the holographic model.
In this study, we shall newly calculate it with a soft-wall AdS/QCD model where the AdS geometry is smoothly cut off at IR by introducing the dilaton field $\Phi(z)$.
The soft-wall model is first considered in Ref.~\cite{Karch:2006pv} based on the bottom-up approach in order to reproduce the correct Regge trajectory of the mass spectrum for excited mesons.
Compared with the hard-wall model which can reproduce the mass of only the ground state hadrons, the soft-wall model is considered to be more suitable to describe the overlap function at the larger bulk coordinate $z'$, because the spectrum of the excited hadrons with large radii contains the information on further long-range (large $z'$ in other words) nonperturbative effects.

We consider the 5D AdS space as the background, 
\begin{equation}
ds^2  = g_{MN} dx^M dx^N  = \frac{1}{{z^2 }}\left( {\eta _{\mu \nu } dx^\mu  dx^\nu   - dz^2 } \right),
\end{equation}
where $\eta _{\mu \nu} = \mbox{diag} (1,-1,-1,-1)$.
$z$ is the fifth (or bulk) coordinate and $\varepsilon \leq z < \infty$ $(\varepsilon \to 0)$.  
The 5D classical action describing a Dirac field is given by
\begin{align}
S_F = &\int d^5 x \sqrt{g}\, e^{-\Phi(z)} \bigg( \frac{i}{2} \bar{\Psi} e^N_A \Gamma^A D_N \Psi	 \nonumber \\
&-\frac{i}{2}(D_N \Psi)^\dagger \Gamma^0 e^N_A \Gamma^A \Psi-(M+\Phi(z))\bar{\Psi}\Psi\bigg), \label{eq:action}
\end{align}
where $e^N_A=z \delta^N_A$, $D_N = \partial_N + \frac{1}{8}\omega_{NAB}[\Gamma^A, \Gamma^B] - iV_N$, and $M$ is the mass of the bulk spinor.
The covariant derivative is introduced to ensure the gauge and diffeomorphism invariance of the action.
In the action, $\Phi (z)$ is the background dilaton field, and we finally choose a solution $\Phi (z) = \kappa^2 z^2$ which reproduces the correct Regge behavior $m_n^2 \propto n$~\cite{Abidin:2009hr}.

The equation of motion for the 5D Dirac field $\Psi$ is written as
\begin{equation}
\bigg[i e^N_A \Gamma^A D_N - \frac{i}{2}(\partial_N\Phi) e^N_A \Gamma^A - (M +\Phi(z)) \bigg] \Psi = 0.
\end{equation}
Subsequently, we define the right-handed and left-handed spinors as $\Psi_{R,L}=(1/2)(1\pm \gamma^5)\Psi$.  
Performing the Fourier transform in the usual 4D space with a four-momentum $p$, 
the 5D spinors can be expressed as
\begin{equation}
\Psi_{R,L} (p,z) = z^\Delta \Psi^0_{R,L} (p) f_{R,L} (p,z),
\end{equation}
where $\Psi^0_{R,L} (p)$ are source fields corresponding to the spin-$1/2$ baryon operators, and $f_{R,L} (p,z)$ are the bulk-to-boundary propagators.  
Taking into account relations  $\!\not\! p \Psi^0_{R(L)} (p) = p \Psi^0_{L(R)} (p)$~(4D Dirac equation) and dropping the vector interaction term, the equation of motion can be rewritten as
\begin{align}
\left(\partial_z - \frac{d/2 + M - \Delta + 2\Phi}{z}\right) f_R &= -pf_L, \nonumber \\
\left(\partial_z - \frac{d/2 - M - \Delta}{z}\right) f_L & = pf_R, \label{eq:EoM}
\end{align}
where $p = \sqrt{p^2}$.

Imposing two boundary conditions, $f_L (p,\varepsilon) = 1$ and the fact that $f_R$ is not singular at $z = \varepsilon$, one can obtain $\Delta = 2-M$.
With a condition $p^2 = m_n^2$, where $m_n$ is a mass of the $n$th Kaluza-Klein (KK) mode, the normalizable modes $\psi _{R,L}^{(n)} (z)$ for the bulk-to-boundary propagators are found as
\begin{align}
&\psi^{(n)}_R(z) = n_R {\xi^{\alpha-1/2}} L^{(\alpha-1)}_n(\xi), \label{eq:psiR} \\
&\psi^{(n)}_L(z) = n_L{\xi^{\alpha}} L^{(\alpha)}_n(\xi), \label{eq:psiL}
\end{align}
where $\xi = \kappa^2 z^2$, $\alpha = M+1/2$,  and $L_n^{(\alpha )} (\xi )$ is the Laguerre polynomial.  The KK mass, $m_n$, is given by the relation $m_n^2 = 4 \kappa^2 (n+\alpha)$.  
Both modes satisfy the normalization conditions,
\begin{equation}
\int dz \frac{e^{-\kappa^2 z^2}}{z^{2M}} \psi^{(n)}_{R,L}\psi^{(m)}_{R,L} = \delta_{nm},
\end{equation}
with
\begin{align}
&n_R = n_L \sqrt{\alpha + n}, \\
&n_L = \frac{1}{\kappa^{\alpha -1}} \sqrt{\frac{2\Gamma (n+1)}{\Gamma (\alpha + n+ 1)}}.
\end{align}

Now we consider the Pomeron~(graviton)-nucleon-nucleon coupling, which can be extracted from the three-point function $\left< 0 | \mathcal{T} \mathcal{O}^i_R(x) T^{\mu\nu} (y) \bar{\mathcal{O}}^j_R(w) | 0 \right>$.
To realize that, we introduce the perturbation to the metric $\eta_{\mu\nu} \to \eta_{\mu\nu} + h_{\mu\nu}$ (then $e^\mu_\alpha \to e^\mu_\alpha - z h^\mu_\alpha /2$), and pick up the $h \bar{\Psi} \Psi$ terms from the action~\cite{Abidin:2009hr}.
Adopting the transverse-traceless gauge, one can obtain the relevant terms as
\begin{equation}
S^{(G)}_F = \int \frac{d^5 x}{z^5} \left( \frac{-iz h_{\mu\nu}}{4} \right) \left( \bar\Psi \Gamma^\mu \tensor{\partial}^\nu \Psi \right).
\end{equation}
Performing the Fourier transform for the fields, the resulting expression is written as
\begin{align}
S^{(G)}_F = &\int \frac{d z}{z^{2M}} e^{-\kappa^2 z^2} \int \frac{d^4 p_2 d^4 q d^4 p_1}{(2\pi )^{12}} \nonumber \\
&\times (2\pi )^4 \delta^4(p_2-q-p_1) \bar\Psi^0_L (p_2) h^0_{\mu\nu} (q) H(q,z) \nonumber \\
&\times \frac{-1}{2} \bigg( f_L(p_2,z)f_L(p_1,z) \gamma^\mu p^\nu \nonumber \\
&+ f_R(p_2,z) f_R(p_1,z) \frac{\!\not\!p_2}{p_2} \gamma^\mu p^\nu \frac{\!\not\!p_1}{p_1} \bigg) \Psi_L(p_1), \label{eq:GA}
\end{align}
where $H(q,z)$ is the bulk-to-boundary propagator.
Comparing the Lorentz structure of Eq.~\eqref{eq:GA} with that of the general parameterization of gravitational form factors,
\begin{align}
\left<p_2,s_2\big| {T}^{\mu\nu}(0) \big| p_1,s_1\right> = &u(p_2,s_2) \left( A(q)\gamma^{(\mu} p^{\nu)} + \cdots \right) \nonumber \\
&\times u(p_1,s_1),
\end{align}
one can uniquely determine $A(q)$.

Since we consider the forward limit to study DIS, we can drop the bulk-to-boundary propagator due to the relation, $H(\varepsilon , z)=1$.
Finally we obtain the nucleon overlap function as
\begin{equation}
P_{24} (z') = \frac{e^{-\kappa^2 z'^2}}{2z'^{2M}}  \left( \psi_R^2 (z') + \psi_L^2 (z') \right).
\end{equation}

For the choice of the model parameters, the authors of Ref.~~\cite{Abidin:2009hr} determined the value of the conformal mass $M=3/2$ by the analysis of the electromagnetic form factors of the nucleon at the large momentum transfer.
In addition, they also fixed the soft-wall parameter $\kappa = 0.350$~[GeV] which could reproduce the masses of the proton and $\rho$ meson simultaneously.
In this paper, we use these values to evaluate the overlap function.

\begin{figure}[bt!]
\includegraphics[width=80mm]{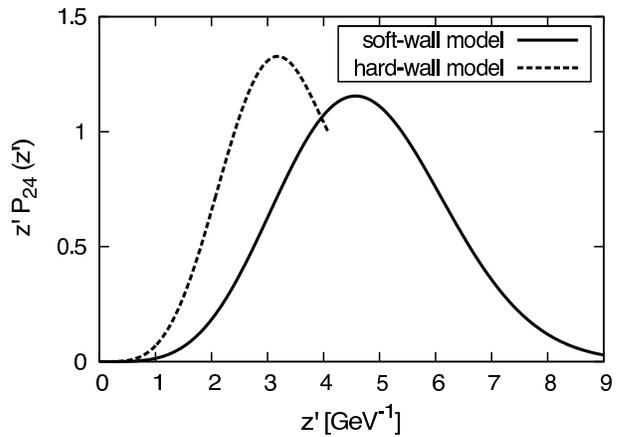}
\caption{
Behavior of the overlap functions for the nucleon in the $z'$ space.
The solid and dashed curves indicate $z'P_{24}(z')$ obtained with the soft-wall model and the hard-wall model~\cite{Watanabe:2012uc}, respectively.
}
\label{fig:OF24}
\end{figure}

We show in Fig.~\ref{fig:OF24}
the $z'$ dependence of $z'P_{24} (z')$, to be compared with that obtained with the hard-wall AdS/QCD model~\cite{Watanabe:2012uc}.
In contrast to the overlap functions for the virtual photon, they are located in the comparatively large $z$ region, $z' \sim 4$~[GeV$^{-1}$] $\sim 0.8$~[fm], which is consistent with the size of the nucleon.
However, the shape of the soft-wall case at the large $z'$ differs significantly from the hard-wall one, which is sharply cut off at the hard-wall parameter, $z' \sim 4.1$~[GeV$^{-1}$].
Namely, details of the nonperturbative effects for $z'  \ge 4$~[GeV$^{-1}$] $\sim 0.8$~[fm] are lost in the hard-wall model.

We emphasize here, as already discussed in Fig.~\ref{fig:r_vs_s}~(b), that the inclusion of the nonperturbative soft-wall effect provides a strong $s$ dependence of the Pomeron kernel if $z'$ is larger than about $4$~[GeV$^{-1}$].  
The overlap function of the nucleon with the soft-wall model is in fact considerably large even for 
$z' > 5$~[GeV$^{-1}$] in Fig.~\ref{fig:OF24}.  
This combined feature of the overlap function and the kernel is indispensable to reproduce the rapid Bjorken-$x$ $(\sim 1/s)$ dependence of the structure function, which will be discussed in the next section.

\section{\label{sec:level4}Numerical results}
To show the numerical results for the structure functions at small $x$, using the obtained overlap functions, we need to fix the three adjustable parameters, $\rho$, $g_0^2$, and $z_0$, which appeared in Eqs.~\eqref{eq:Fic} and \eqref{eq:kernel_mod}.  
$\rho$ plays the role of determining the energy dependence of the cross section, and $g_0^2$ controls the magnitude of the structure function.  
The parameter  $z_0$ controls the strength of the conformal breaking effect.
We expected $1/z_0$ to be the same order of the QCD scale parameter, i.e. $1/z_0 \sim \Lambda_{QCD}$.
As a result, we find $\rho = 0.856$ and $g_0^2 = 0.908 \times 10^2$ for the conformal kernel case, and $\rho = 0.783$, $g_0^2 = 1.62 \times 10^2$, and $z_0 = 4.25$~[GeV$^{-1}$] for the modified kernel case to reproduce the experimental data for the proton $F_2$ structure function measured at HERA~\cite{Aaron:2009aa}.

\begin{figure*}[bt!]
\includegraphics[width=0.97\textwidth]{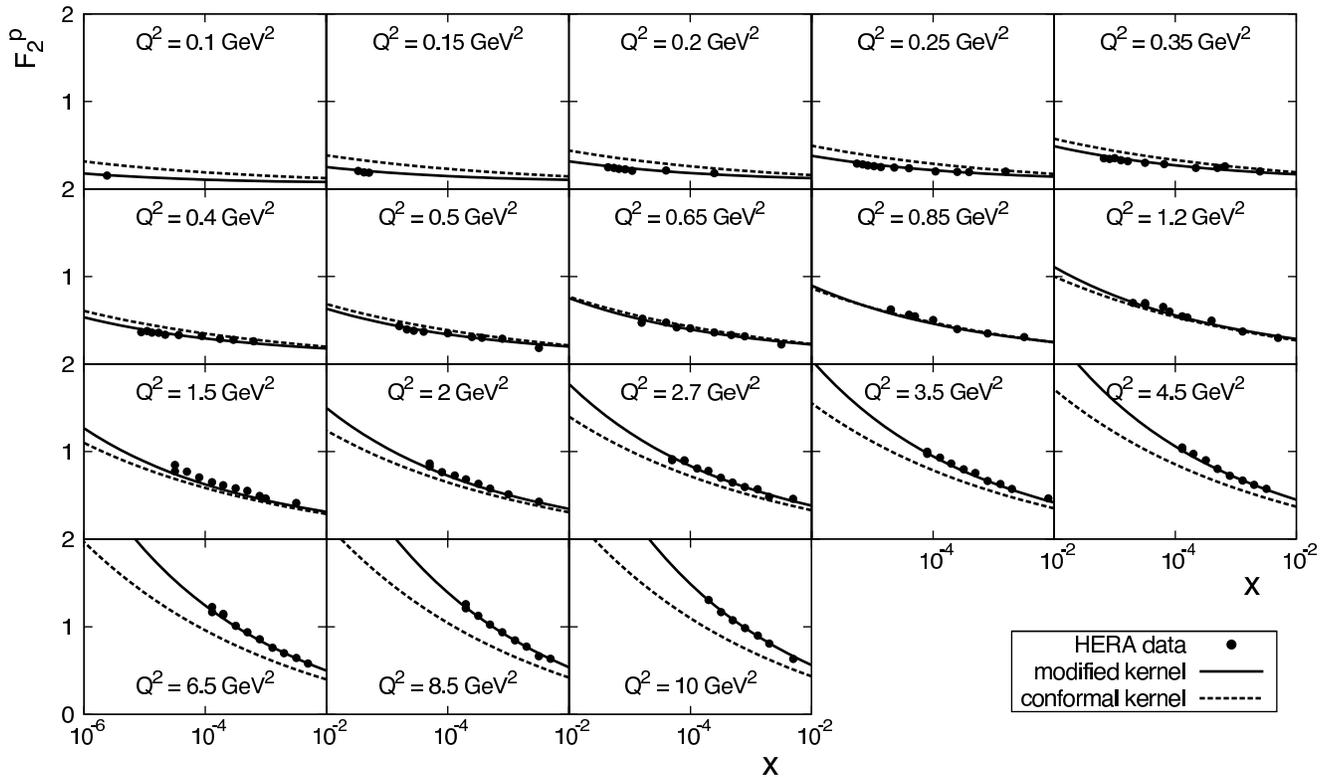}
\caption{
$F_2^p(x,Q^2)$ as a function of the Bjorken-$x$ for various $Q^2$.
In each panel, the solid and dashed curves represent our calculations with the modified and conformal kernels, respectively.  HERA data~\cite{Aaron:2009aa} are depicted by circles.
}
\label{fig:F2_N}
\end{figure*}

\begin{figure}[bt!]
\includegraphics[width=83mm]{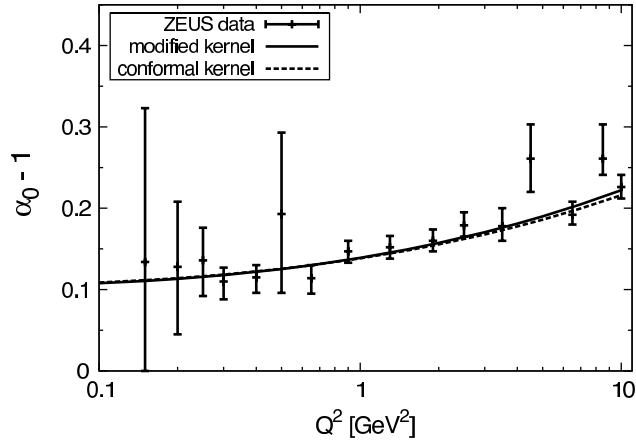}
\caption{
$Q^2$ dependence of the effective Pomeron intercept $\alpha _0(Q^2)-1$.
The solid and dashed curves represent our calculations with the modified and conformal kernels, respectively.
HERA data~\cite{Breitweg:1998dz} are depicted with error bars.
}
\label{fig:interceptN}
\end{figure}

First we show in Fig.~\ref{fig:F2_N}
the resulting $F_2^p (x,Q^2)$ as a function of the Bjorken-$x$ for various $Q^2$.
From the figure, one can see that the calculations with the modified kernel are in good agreement with the HERA data, while the $Q^2$ dependence obtained with the conformal kernel is too weak to reproduce the data.
With these calculations, we extract the scale dependence of the effective Pomeron intercept $\alpha_0 (Q^2)-1$ shown in Fig.~\ref{fig:interceptN}, assuming the form of Eq.~\eqref{eq:F2withPomeron}.
In both conformal and modified cases, one can see the rise of $\alpha _0(Q^2)-1$ with $Q^2$, which are reasonably consistent with the experimental data.

\begin{figure}[bt!]
\includegraphics[width=79mm]{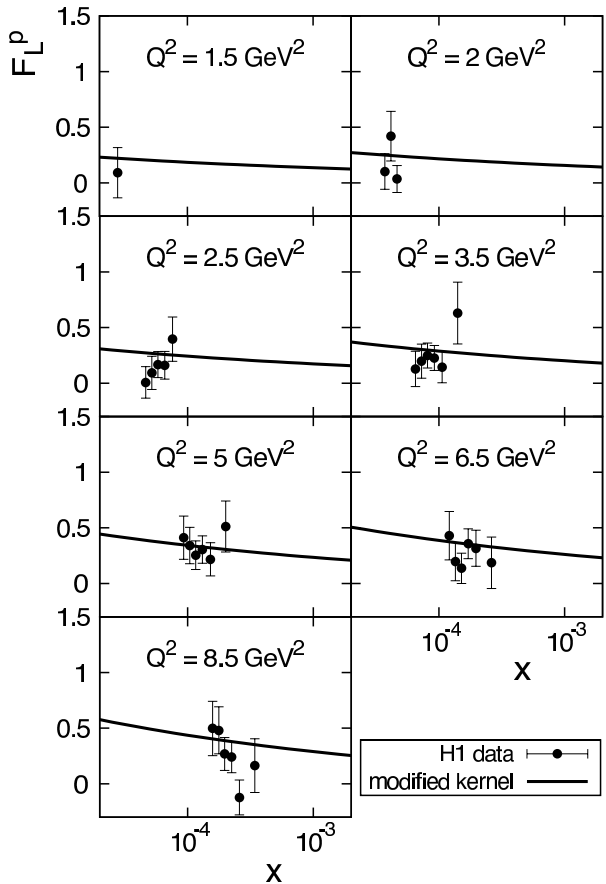}
\caption{
$F_L^p(x,Q^2)$ as a function of the Bjorken-$x$ for various $Q^2$.
In each panel, the solid curves represent our calculations with the modified kernel.
HERA data~\cite{Collaboration:2010ry} are depicted by circles with error bars.
}%
\label{fig:FL_N}%
\end{figure}

Hereafter, we adopt the modified kernel and consider the longitudinal structure function $F_L^p$.  
Since the parameters are fixed in the calculations for the $F_2$ structure function, we can evaluate $F_L$ replacing $P_{13}^{(2)}$ with $P_{13}^{(L)}$ in Eq.~\eqref{eq:Fic}.
We show in Fig.~\ref{fig:FL_N}
the resulting $F_L^p(x,Q^2)$ as a function of the Bjorken-$x$ for several $Q^2$.  
Although the experimental data recently published by the H1 Collaboration~\cite{Collaboration:2010ry} have comparatively large error bars, our calculations are in good agreement with the data.

\begin{figure}[bt!]
\includegraphics[width=81mm]{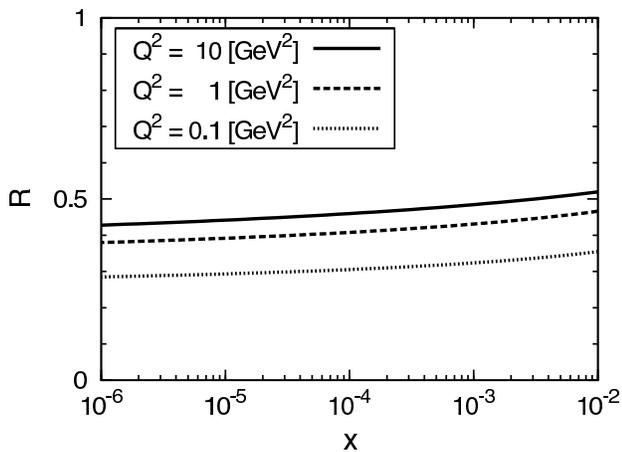}
\caption{
The solid, dashed, and dotted curves represent the Bjorken-$x$ dependence of the resulting longitudinal-to-transverse ratio $R(x,Q^2)=F_L^p(x,Q^2)/F_T^p(x,Q^2)$ for $Q^2 = 10$~[GeV$^2$], $1$~[GeV$^2$], and  $0.1$~[GeV$^2$], respectively.
}
\label{fig:Rvsx}
\end{figure}

Subsequently, we consider the longitudinal-to-transverse ratio of the structure functions, $R(x,Q^2)=F_L^p(x,Q^2)/F_T^p(x,Q^2)$ where $F_T = F_2 - F_L$, at small $x$.
In Fig.~\ref{fig:Rvsx},
we show the Bjorken-$x$ dependence of the ratio for $Q^2 = 10$~[GeV$^2$], $1$~[GeV$^2$], and  $0.1$~[GeV$^2$].  
From the figure, one can see a non-trivial $x$ and $Q^2$ dependence of $R$.  
Our calculations show that the ratio $R$ increases with the Bjorken-$x$ and/or the scale $Q^2$ increasing.
The $x$ dependence is comparatively weak, but the $Q^2$ dependence is obviously substantial.
Qualitative behavior of our results is similar to that of the DGLAP fit in the ACOT scheme in Ref.~\cite{Collaboration:2010ry}.

\section{\label{sec:level5}Summary and discussion}
In this paper, we have studied the nucleon $F_2$ and $F_L$ structure functions focusing on the typical nonperturbative region, where $10^{-6} \leq x \leq 10^{-2}$ and $0.1 \leq Q^2 \leq 10$~[GeV$^2$], in the framework of the holographic QCD.
In particular, we have improved the description of the target nucleon and presented a more consistent model setup as compared with the previous study.
To obtain the Pomeron-nucleon coupling in the AdS space, we have used a soft-wall AdS/QCD model of the nucleon, in which the AdS geometry is smoothly cut off at IR.
In contrast to the hard-wall model, the soft-wall model can reproduce the correct Regge behavior of the mass spectrum for both the ground state and the excited states of hadrons.
This may indicate the soft-wall model could provide a better description for nonperturbative properties of the nucleon.
In fact, the behavior of the obtained overlap function of the nucleon in the AdS space is apparently different from that calculated via the hard-wall model in the previous study.

In our framework, once the two (the conformal kernel case) or three (the modified kernel case) adjustable parameters are fixed with the experimental data, one can calculate not only the $F_2$ structure function but also the longitudinal structure function $F_L$ without any additional parameter.  
Our calculations for $F_2^p (x,Q^2)$ with the modified kernel are in quite good agreement with the experimental data.

The results for $F_L^p (x,Q^2)$, which have been newly calculated in this study, are also consistent with the recent HERA data.
We have also considered the longitudinal-to-transverse ratio $R$, and found the resulting ratio only slightly increases as the  Bjorken-$x$ increases.
In particular, the ratio increases by about $30\%$ from $Q^2 = 0.1$~[GeV$^2$] to $1$~[GeV$^2$], which is due to the purely nonperturbative effect.

Here, we comment on the realization of the Callan-Gross (CG) relation~\cite{Callan:1969uq}, $F_2(x) = 2 x F_1(x)$, in the holographic description of the nucleon structure function.
This relation exactly holds in the free quark-parton approximation.
In the first work on DIS in Ref.~\cite{Polchinski:2002jw}, the rigorous supergravity calculation provides a relation $F_2 = 2 F_1$ (see section 4 of Ref.~\cite{Polchinski:2002jw}), which violates the CG relation.
To obtain it, they considered only the supergravity states as the intermediate states that were produced after the initial hadron absorbs the virtual photon.
On the other hand, they also considered the structure function at the small $x$ region taking into account the intermediate states with excited strings (section 5 of Ref.~\cite{Polchinski:2002jw}).
These results show the CG relation approximately holds with slight modifications.
The calculations on the BPST Pomeron, which is adopted in this work, was done in a similar way to describe the reggeized graviton exchange as the Pomeron.
The size of the deviation from the  CG relation can be seen in the longitudinal-to-transverse ratio, $R \simeq ( F_2 - 2 x F_1 ) / ( 2 x F_1 )$.
Our results, $R \sim 0.3 - 0.5$, in Fig.~\ref{fig:Rvsx} indicate obvious deviations from the CG relation due to the nonperturbative effect.
Further studies are needed from both theoretical and experimental points of view to understand the origins of such deviations in the nonperturbative kinematical region.

Finally, we make two concluding remarks.
First, we have confirmed that our framework works well even if we switch to the soft-wall AdS/QCD model from the hard-wall model, although the calculated Pomeron-nucleon coupling is clearly different from the previous hard-wall calculations.
It seems, however, that such differences are absorbed into the readjustment of some of the model parameters, yielding quantitatively very similar numerical results for both soft- and hard-wall cases.
Therefore, the hard-wall model is still a valuable practical tool when we build a model of the ground state hadrons.
Second, the holographic approach provides a reasonable description of the high energy scattering phenomena, and further studies are certainly needed.
The AdS/QCD can be a powerful analytical tool as a ``building block'' to build a model of the nonperturbative QCD, although this approach involves fundamental  uncertainties related to the original AdS/CFT correspondence.
In principle, by giving the appropriate density distributions in the AdS space for the incident and target particles, other two-body high energy scattering processes can be studied  in this framework.
Some applications are under consideration.

\begin{acknowledgments}
A.W. acknowledges Chung-Wen Kao, Yoshio~Kitadono, and Chung-I Tan for valuable comments and discussion.
We are grateful to Hsiang-nan~Li for a careful reading of the manuscript.
The work of A.W. was supported by the National Science Council of R.O.C. under Grant No. NSC-102-2811-M-001-024.
\end{acknowledgments}

\end{document}